\documentclass[aps,prl,reprint,superscriptaddress]{revtex4-2}

\usepackage{graphicx}
\graphicspath{figures/}
\usepackage{dcolumn}
\usepackage{amssymb}
\usepackage{amsmath}
\usepackage{fancyhdr}
\usepackage{bm}
\usepackage{times}
\usepackage{mhchem}
\usepackage{ulem}
\usepackage{lineno}
\usepackage[colorlinks,linkcolor=blue,anchorcolor=blue,citecolor=blue,urlcolor=blue]{hyperref}

\usepackage{changes}

\UseRawInputEncoding
\begin{document}

\title{Magnon-Skyrmion Hybrid Quantum Systems: Tailoring Interactions via Magnons}
\author{Xue-Feng Pan}
\author{Peng-Bo Li}
\email{lipengbo@mail.xjtu.edu.cn}
\author{Xin-Lei Hei}
\affiliation{Ministry of Education Key Laboratory for Nonequilibrium Synthesis and Modulation of Condensed Matter, Shaanxi Province Key Laboratory of Quantum Information and Quantum Optoelectronic Devices, School of Physics, Xi'an Jiaotong University, Xi'an 710049, China}
\author{Xichao Zhang}
\author{Masahito Mochizuki}
\affiliation{Department of Applied Physics, Waseda University, Okubo, Shinjuku-ku, Tokyo 169-8555, Japan}
\author{Fu-Li Li}
\affiliation{Ministry of Education Key Laboratory for Nonequilibrium Synthesis and Modulation of Condensed Matter, Shaanxi Province Key Laboratory of Quantum Information and Quantum Optoelectronic Devices, School of Physics, Xi'an Jiaotong University, Xi'an 710049, China}
\author{Franco Nori}
\affiliation{Theoretical Quantum Physics Laboratory, Cluster for Pioneering Research, RIKEN, Wakoshi, Saitama 351-0198, Japan}
\affiliation{Center for Quantum Computing, RIKEN, Wakoshi, Saitama 351-0198, Japan}
\affiliation{Physics Department, The University of Michigan, Ann Arbor, Michigan 48109-1040, USA }

\date{\today}

\begin{abstract}
Coherent and dissipative interactions between different quantum systems are essential for the construction of hybrid quantum systems and the investigation of novel quantum phenomena. Here, we propose and analyze a magnon-skyrmion hybrid quantum system, consisting of a micromagnet and nearby magnetic skyrmions. We predict a strong coupling mechanism between the magnonic mode of the  micromagnet and the quantized helicity degree of freedom of the skyrmion. We show that with this hybrid setup it is possible to induce magnon-mediated \textit{nonreciprocal interactions and responses} between  distant skyrmion qubits or between skyrmion qubits and other quantum systems like superconducting qubits. This work provides a quantum platform for the investigation of diverse quantum effects and quantum information processing with magnetic microstructures.
\end{abstract}
\maketitle

\textit{Introduction.}---Coherent and dissipative couplings between degrees of freedom of completely different nature are the fundamental issue of quantum science and technology. These diverse types of interactions are the foundation for quantum information processing with hybrid quantum systems~\cite{2013XiangP623653,2021HarderP201101201101,2006GaebelP408413,2012VerhagenP6367,2013BraakmanP432437,2015ViennotP408411,2017AstnerP140502140502} and have been widely used to explore new quantum phenomena like nonreciprocal transport~\cite{2015MetelmannP2102521025,2019WangP127202127202} and non-Hermitian physics~\cite{2020CaoP3040130401,2022ZhangP2403824038}. Quantum magnonics~\cite{2015ChumakP453461,2021LiP164,2022YuanP174,2022ZareRameshtiP161,2019LachanceQuirionP7010170101,2016TabuchiP729739,2014KrawczykP123202123202,2021PirroP11141135,2021YuP159}, with the use of magnons (spin-wave quanta), provides a promising platform to study different types of quantum interactions. In particular, the magnon mode in the ferromagnetic material yttrium iron garnet (YIG) with the advantage of high spin density and low damping rate has attracted much attention ~\cite{2018WangP5720257202,2022PanP5442554425,2019LiuP134421134421,2020YuanP100402100402,
2020XieP4233142331,2017ZhangP13681368,2019CaoP214415214415,2019ZhangP5440454404,2011WangP146602146602,2018QinP54455445}. Magnons, similar to phonons and photons~\cite{2012PootP273335,2019WhiteleyP490495,2020ClerkP257267,2021PerdriatP,2021BlaisP2500525005,2022ShandilyaP75387571,2017GuP1102,2019FriskKockumP1940}, can strongly couple to various quantum systems~\cite{2015TabuchiP405408,2022KounalakisP3720537205,2020NeumanP247702247702,2021HeiP4370643706,2023HeiP7360273602,2022XiongP245310245310,2017AndrichP2828,2020CandidoP1100111001,2021FukamiP4031440314,2021SkogvollP6400864008,2022GonzalezBallesteroP7541075410,2024BejaranoP20422042,2010SoykalP7720277202,2015LambertP5391053910,2015ZhangP1501415014,2016KostylevP6240262402,2017SharmaP9441294412,2019BhoiP134426134426,2024XuP116701116701,2016ZhangP15012861501286,2018LiP203601203601,2022KaniP1360213602,2022ShenP243601243601,2023AsjadP37,2020ColombanoP147201147201,2021LiP}. Magnon-based hybrid quantum systems (such as magnon-photon~\cite{2010SoykalP7720277202,2015LambertP5391053910,2015ZhangP1501415014,2016KostylevP6240262402,2017SharmaP9441294412,2019BhoiP134426134426,2024XuP116701116701}, magnon-phonon~\cite{2016ZhangP15012861501286,2018LiP203601203601,2022KaniP1360213602,2022ShenP243601243601,2023AsjadP37,2021LiP,2020ColombanoP147201147201}, magnon-solid state spin~\cite{2020NeumanP247702247702,2021HeiP4370643706,2022XiongP245310245310,2023HeiP7360273602,2017AndrichP2828,2020CandidoP1100111001,2021FukamiP4031440314,2021SkogvollP6400864008,2022GonzalezBallesteroP7541075410,2024BejaranoP20422042}, and magnon-superconducting qubit (SQ) hybrid setups~\cite{2022KounalakisP3720537205,2020WolskiP117701117701}), have been proposed and investigated.  A plethora of interesting quantum effects have been investigated in such magnon-based quantum setups, which include the generation of non-classical states~\cite{2023XuP193603193603,2022KounalakisP3720537205}, ground state cooling~\cite{2022KaniP1360213602,2023AsjadP37}, quantum transducers~\cite{2023HeiP7360273602}, multi-body entanglement~\cite{2018LiP203601203601,2023HeiP7360273602}, and quantum state conversion~\cite{2020NeumanP247702247702,2021FukamiP4031440314} . All these make quantum magnons particularly attractive for quantum technologies, and it is highly appealing to explore novel systems and novel ways of manipulating and
coupling the magnetization to different degrees of freedom.

Recently, magnetic nanostructures such as skyrmions~\cite{2017FertP1703117031,2018EverschorSitteP240901240901,2019OchoaP19300051930005,2020BackP363001363001,2020LonskyP100903100903,2022ReichhardtP3500535005} have attracted great interests in the field of quantum science and technology~\cite{2012MochizukiP1760117601,2013LinP6040460404,2014SchuetteP9442394423,2015RoldanMolinaP245436245436,
2015ZhangP102401102401,2017PsaroudakiP4104541045,2018PsaroudakiP237203237203,
2019CasiraghiP145145,2020CapicP415803415803,2020HirosawaP207204207204,2021KhanP100402100402,
2021LiensbergerP100415100415,2022HirosawaP4032140321}.
In frustrated magnets, the skyrmion has an internal degree of freedom associated with the rotation of the helicity~\cite{2015LeonovP82758275,2016LinP6443064430,2019KurumajiP914918,2017ZhangP17171717, 2020YaoP8303283032, 2023XiaP106701106701}, and by quantizing the collective helicity coordinate, two categories of qubits can be established~\cite{2021PsaroudakiP6720167201}, providing a promising platform for carrying out quantum computation. However,
the coupling of this quantized helicity degree of freedom in skyrmions to other quantum systems like magnons remains largely unexplored. The investigation of coupling skyrmion qubits to other quantum degrees of freedom has its own significance and interest: First, this could enable the scalability of skyrmion qubits, since long range coherent interactions between distant skyrmions could be implemented by using the coupled quantum system as an intermediary. Second, this may allow to study  novel quantum effects such as
nonreciprocal quantum phenomena through the use of other useful methods like quantum reservoir engineering.

In this work, we propose and analyze a hybrid quantum system composed of a YIG micromagnet and a skyrmion, predicting that strong coupling between the magnon mode of the micromagnet and the quantized helicity degree of freedom of the skyrmion is feasible. Here, the micromagnet acts as a microwave nanomagnonic cavity to concentrate the magnonic excitations, while the skyrmion qubit behaves very similar to a superconducting charge qubit. We find that the coherent coupling between the magnonic excitation and the skyrmion qubit is well described by the Jaynes-Cummings (JC) model~\cite{1963JaynesP89109,1993ShoreP11951238}. To further enhance the coupling strength, we take into account the anisotropy of the YIG micromagnet, which results in the magnon-Kerr effect and allows the magnons to be squeezed~\cite{2016WangP224410224410,2019KongP3400134001,2022XiongP245310245310,2023JiP180409180409} in analogy to squeezed phonons~\cite{2016LemondeP1133811338,2019GeP3050130501,2020LiP153602153602,2021BurdP898902} and photons~\cite{2015LueP9360293602,2018QinP9360193601,2018LerouxP9360293602,2020GroszkowskiP203601203601,2021ChenP2360223602}. Therefore, the coupling strength can be increased exponentially. By employing the magnons as  an intermediary, it is possible to create tunable coherent couplings as well as dissipative couplings between distant skyrmion qubits or between skyrmion qubits and superconducting qubits. This allows for \textit{nonreciprocal interactions and responses} between distant qubits via using the magnon-mediated dissipative coupling. For YIG micromagnets, the geometry is unspecified, which can be spherical, thin-film or even bulk. The hybrid quantum systems proposed in this work give an all-magnetic platform and greatly broaden the avenue for quantum information processing.

\begin{figure}
    \centering
    \includegraphics[width=0.48\textwidth]{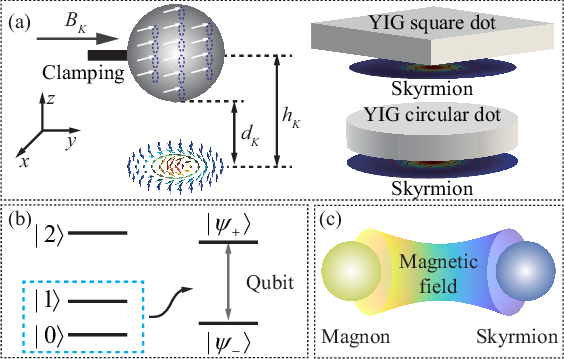}
    \caption{(a) Hybrid quantum system consisting of a YIG sphere (a YIG square or circular dot) and a skyrmion. (b) Energy level structure of skyrmion qubits. (c) Coupling mechanisms of magnons and skyrmion qubits.}
    \label{FIG1}
\end{figure}

\textit{The setup.}---As illustrated in Fig.~\ref{FIG1}(a), we consider a hybrid quantum system consisting of a micromagnet and a skyrmion, where the skyrmion is located beneath the micromagnet. The micromagnet can adopt various shapes such as spheres, square or circular  dots~\cite{2020DingP1401714017,2020WangP144428144428,2022ChaudhuriP1202212029,2023SrivastavaP6407864078,2023MerboucheP230616094}, and other configurations. Our primary focus, here, is on a hybrid system that integrates a YIG sphere with a skyrmion [the left half of Fig.~\ref{FIG1}(a)]. The vertical distances from the skyrmion to the YIG sphere's surface and center are $d_K$ and $h_K$, respectively. In the YIG sphere, long-lived spin wave modes can be excited by applying a uniform bias magnetic field $B_K$~\cite{2020GonzalezBallesteroP125404125404,2020GonzalezBallesteroP9360293602}. Here, we solely take into account the Kittel mode, all spins in the micromagnet precessing in phase and with the same amplitude. The Kittel mode's free Hamiltonian can then be expressed as $\hat{H}_K=\omega_K\hat{s}_K^{\dagger}\hat{s}_K$, where $\hat{s}_K$ ($\hat{s}_K^{\dagger}$) represents the annihilation (creation) operator and $\omega_K=\gamma_e B_K$ denotes the resonance frequency (setting $\hslash=1$)~\cite{SM}. The multilayer structure~\cite{2013SunP167201167201,2015GilbertP84628462}, which includes square or circular  dots and skyrmions as illustrated in the right half of Fig.~\ref{FIG1}(a), is discussed in detail in Ref.~\cite{SM}.

A skyrmion is a non-collinear spin texture with a centrosymmetric spiraling structure, as shown in  Fig.~S1(a). The skyrmion is located at the center of the coordinate, and its spin at position $\widetilde{\boldsymbol{r}}=(\widetilde{x},\widetilde{y})=(\widetilde{\rho},\phi)$ is $\boldsymbol{S}$, corresponding to the magnetic moment $\boldsymbol{\mu}_s=-g\mu_B\boldsymbol{S}$ with Land\'e factor $g$ and Bohr magneton $\mu_B$, where $(\widetilde{x},\widetilde{y})$ and $(\widetilde{\rho},\phi)$ denote Cartesian and Polar coordinates, respectively. For a typical skyrmion, the spin direction is given by the normalized spin $\boldsymbol{s}=\boldsymbol{S}/\vert \boldsymbol{S}\vert=(s_x,s_y,s_z)=[\sin \Theta(\widetilde{\rho}) \cos \Phi,\sin \Theta(\widetilde{\rho}) \sin \Phi,\cos \Theta(\widetilde{\rho})]$, with $\Phi=\phi+\varphi_0$ and helicity $\varphi_0$. The helicity $\varphi_0$ is the internal degree of freedom of skyrmions in frustrated magnets. The $z$ component $s_z$ of the normalized spin $\boldsymbol{s}$ is depicted in Fig.~S1(d), with the central spin oriented downward and the edge spin oriented upward. Skyrmion-based $\mathfrak{S}_z$ qubits can be constructed by  quantizing the helicity $\varphi_0$. Utilizing the collective coordinate quantization technique~\cite{SM,1975GoldstoneP14861498,1994DoreyP35983611,2017PsaroudakiP4104541045,2021PsaroudakiP6720167201,2022PsaroudakiP104422104422}, the Hamiltonian of $\mathfrak{S}_z$ qubits can be represented as $\hat{\mathcal{H}}_{\mathfrak{S}_z}=\bar{\bar{\kappa}}_z\hat{\mathfrak{S}}_z^2-\bar{\bar{h}}_z\hat{\mathfrak{S}}_z-\bar{\bar{\varepsilon}}_z\cos\hat{\varphi}_0$, with the collective coordinate $\hat{\varphi}_0$ and its conjugate momentum $\hat{\mathfrak{S}}_z$, where the parameters $\bar{\bar{\kappa}}_z$, $\bar{\bar{h}}_z$, and $\bar{\bar{\varepsilon}}_z$ are defined in detail in Ref.~\cite{SM}. As shown in Fig.~\ref{FIG1}(b), the energy levels ($\{\vert0\rangle,\vert1\rangle,\vert2\rangle,\cdots\}$) of $\mathfrak{S}_z$ qubits are anharmonic ~\cite{2021PsaroudakiP6720167201,2022PsaroudakiP104422104422}, the non-harmonicity of which is generally larger than $20\%$. Hence, the Hamiltonian $\hat{\mathcal{H}}_{\mathfrak{S}_z}$ in the subspace $\{\vert0\rangle,\vert1\rangle\}$ is given by $\hat{H}_{\rm{Sky}}=\mathcal{A}_0/2\hat{\sigma}_z^{\rm{sub}}-\mathcal{B}_0/2\hat{\sigma}_x^{\rm{sub}}$, where $\mathcal{A}_0\equiv\bar{\bar{\kappa}}_z-\bar{\bar{h}}_z$, $\mathcal{B}_0\equiv\bar{\bar{\varepsilon}}_z$, and Pauli operators $\hat{\sigma}_z^{\rm{sub}}\equiv\vert1\rangle\langle1\vert-\vert0\rangle\langle0\vert$ and $\hat{\sigma}_x^{\rm{sub}}\equiv\vert1\rangle\langle0\vert+\vert0\rangle\langle1\vert$.

\textit{Interactions between the magnon and the skyrmion.}---Magnons and skyrmion qubits are coupled via magnetic fields [Fig.~\ref{FIG1}(c)]. At position $\boldsymbol{r}$, a YIG sphere with magnetic moment $\boldsymbol{\mu}=\boldsymbol{M}4\pi R_K^3/3$, radius $R_K$, and magnetization $\boldsymbol{M}$ produces a magnetic field that is described by $\boldsymbol{B}=\frac{\mu_0}{4\pi}\left[\frac{3\boldsymbol{r}(\boldsymbol{\mu}\cdot\boldsymbol{r})}{r^5}-\frac{\boldsymbol{\mu}}{r^3}\right]$~\cite{2020GieselerP163604163604}, where $\mu_0$ is the vacuum permeability and $\boldsymbol{r}=(\rho\cos\phi,\rho\sin\phi,-h_K)$ is the position vector from the magnetic sphere's center to any point on the skyrmion plane. By introducing the quantized magnetization operator $\hat{\boldsymbol{M}}=M_K[\widetilde{\boldsymbol{m}}_K\hat{s}_K+\widetilde{\boldsymbol{m}}_K^*\hat{s}_K^{\dagger}]$, with the Kittel mode function $\widetilde{\boldsymbol{m}}_K=\hat{\boldsymbol{e}}_z+i\hat{\boldsymbol{e}}_x$~\cite{2020GonzalezBallesteroP125404125404,2020GonzalezBallesteroP9360293602,2021HeiP4370643706}, the quantized magnetic field generated by the YIG sphere can be expressed as $\hat{\boldsymbol{B}}=(\hat{B}_x,\hat{B}_y,\hat{B}_z)$ (the detailed definition can be found in Ref.~\cite{SM}). \textit{The skyrmion-magnon interaction} is described by $\hat{H}_{\rm{KS}}=-g\mu_B\bar{S}/a^2\int d\widetilde{\boldsymbol{r}}\hat{\boldsymbol{B}}\cdot\boldsymbol{s}$~\cite{2021PsaroudakiP6720167201,SM}, with $\bar{S}$ the effective spin. By utilizing the collective operators $\hat{\varphi}_0$ and $\hat{\mathfrak{S}}_z$ and expanding $\hat{H}_{\rm{KS}}$ in the subspace $\{\vert0\rangle,\vert1\rangle\}$, the interaction Hamiltonian is reduced to
    \begin{equation}
        \color[rgb]{0,0,0}
        \hat{H}_{\rm{KS}}=\frac{\lambda_{\rm{KS}}^{xy}}{2}\left(\hat{s}_K+\hat{s}_K^{\dagger}\right)\hat{\sigma}_x^{\rm{sub}}+\frac{\lambda_{\rm{KS}}}{2}\left(\hat{s}_K+\hat{s}_K^{\dagger}\right)\hat{\sigma}_z^{\rm{sub}},
        \label{HKSQuantum}
    \end{equation}
where the skyrmion-magnon coupling strength is given by~\cite{SM}
    \begin{equation}
        \lambda_{\rm{KS}}=\frac{2\pi g\mu_B\bar{S}\mu_0R_K^3 M_K}{3a^3\Lambda}\mathcal{F}(\rho)
    \end{equation}
with zero-point magnetization $M_K=\sqrt{\hslash\gamma_eM_s/(2V_K)}$, gyromagnetic ratio $\gamma_e$, saturation magnetization $M_s$, volume of the YIG sphere $V_K$, and $\Lambda=\int d\boldsymbol{r}(1-\cos\Theta_0)$. $\mathcal{F}(\rho)$ is a dimensionless integral~\cite{SM}. {\color[rgb]{0,0,0}The transverse coupling strength $\lambda_{\rm{KS}}^{xy}$ is presented in detail in Ref.~\cite{SM}}.

The total Hamiltonian of the skyrmion-magnon hybrid quantum system is $\hat{H}_{\rm{TKS}}=\hat{H}_K+\hat{H}_{\rm{Sky}}+\hat{H}_{\rm{KS}}$. The eigenvectors of the $\mathfrak{S}_z$ qubit Hamiltonian $\hat{H}_{\rm{Sky}}$ are given by $\vert \psi _+\rangle=\cos\theta\vert1\rangle - \sin\theta \vert0\rangle$ and $\vert \psi _-\rangle=\sin\theta\vert1\rangle + \cos\theta \vert0\rangle$, with their corresponding eigenenergies $\mathcal{E}_\pm=\pm1/2\sqrt{\mathcal{A}_0^2+\mathcal{B}_0^2}$ and $\tan (2\theta)=\mathcal{B}_0/\mathcal{A}_0$ [Fig.~\ref{FIG1}(b)]. In the subspace $\{\vert \psi _+\rangle,\vert \psi _-\rangle\}$, the interaction Hamiltonian $\hat{H}_{\rm{KS}}$ can be expanded as $\hat{H}_{\rm{KS}}=\lambda_{\rm{KS}}^{xy}/2(\hat{s}_K+\hat{s}_K^{\dagger})[-\sin(2\theta)\hat{\sigma}_z+\cos(2\theta)\hat{\sigma}_x]+\lambda_{\rm{KS}}/2(\hat{s}_K+\hat{s}_K^{\dagger})[\cos(2\theta)\hat{\sigma}_z+\sin(2\theta)\hat{\sigma}_x]$, with Pauli operators $\hat{\sigma}_z=\vert\psi_+\rangle\langle\psi_+\vert-\vert\psi_-\rangle\langle\psi_-\vert$, $\hat{\sigma}_+=\vert\psi_+\rangle\langle\psi_-\vert$, and $\hat{\sigma}_-=\vert\psi_-\rangle\langle\psi_+\vert$. When the $\mathfrak{S}_z$ qubit works in the vicinity of the degeneracy point, $\sin(2\theta)\sim 1$ and $\cos(2\theta)\sim 0$ can be obtained~\cite{SM}. Then, using the rotating-wave approximation the hybrid quantum system Hamiltonian becomes
    \begin{equation}
        \hat{H}_{\rm{TKS}}=\frac{\omega_q}{2}\hat{\sigma}_z+\omega_K\hat{s}_K^{\dagger}\hat{s}_K+\bar{\lambda}_{\rm{KS}}\left(\hat{s}_K\hat{\sigma}_++\hat{s}_K^{\dagger}\hat{\sigma}_-\right),
        \label{HTKS}
    \end{equation}
where $\omega_q=\mathcal{E}_+ - \mathcal{E}_-$ denotes the resonant frequency of the qubit and the coupling strength is written as $\bar{\lambda}_{\rm{KS}}=\lambda_{\rm{KS}}\sin(2\theta)/2$. Note that the term $\lambda_{\rm{KS}}^{xy}\sin(2\theta)/2(\hat{s}_K+\hat{s}_K^{\dagger})\hat{\sigma}_z$ has been neglected under the rotating-wave approximation, which is discussed in detail in Ref.~\cite{SM}.

\begin{figure}
    \centering
    \includegraphics[width=0.48\textwidth]{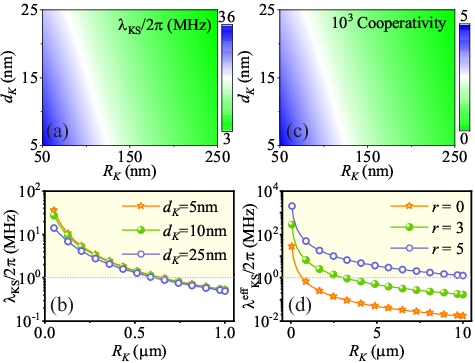}
    \caption{Contour maps of the coupling strength $\lambda_{\rm{KS}}$ versus $R_K$ and $d_K$ are visualized in (a). (b) Coupling strength $\lambda_{\rm{KS}}$ as a function of $R_K$.  (c) Contour maps of the cooperativity $\mathcal{C}$. The dissipation used to calculate the cooperativity is $\gamma_K/2\pi=\gamma_{\rm{Sky}}/2\pi=1~{\rm{MHz}}$. (d) The enhanced coupling strength by the two-magnon drive ($d_K=10~{\rm{nm}}$).}
    \label{FIG2}
\end{figure}
As shown in Figs.~\ref{FIG2}(a, b), the coupling strength $\lambda_{\rm{KS}}$ is plotted as a function of the radius $R_K$ and distance $d_K$. According to contour maps [Fig.~\ref{FIG2}(a)], the coupling strength can reach $20~\rm{MHz}$ if the YIG sphere's radius can be reduced to under $100~\rm{nm}$.  In Fig.~\ref{FIG2}(b), taking different $d_K$, the coupling strength $\lambda_{\rm{KS}}$ decreases with increasing $R_K$, and the shaded region depicts the strong-coupling region. Here, we have assumed $d_K=10~\rm{nm}$. $\lambda_{\rm{KS}}/2\pi$ can reach $12.7~\rm{MHz}$ and $5.2~\rm{MHz}$, corresponding to $R_K=100~\rm{nm}$ and $R_K=200~\rm{nm}$, respectively.  To quantitatively analyze the system's quantum effects, we introduce the cooperativity $\mathcal{C}=4\lambda_{\rm{KS}}^2/(\gamma_{K}\gamma_{\rm{Sky}})$~\cite{2020ClerkP257267}, where $\gamma_{K}$ and $\gamma_{\rm{Sky}}$ represent the dissipation of the magnon and the skyrmion, respectively. Figure~\ref{FIG2}(c) indicates that the system can reach the strong-coupling regime ($\mathcal{C}>1$) in a broad range of $d_K$ and $R_K$.

\textit{Exponentially enhanced coupling strength.}---As shown in Fig.~\ref{FIG2}(b), the radius of the YIG sphere that achieves strong coupling is $0.6~{\rm{\mu m}}$. To ensure that the large-size YIG sphere still achieves strong coupling, we utilize the parametric amplification technique to enhance the coupling strength exponentially. Here, we take into account the YIG sphere's anisotropic energy, which results in the magnon-Kerr effect~\cite{2016WangP224410224410,2019KongP3400134001}. A microwave drive is used to enhance the Kerr effect of the YIG sphere, which is described by $\hat{H}_d=\Omega_d(\hat{s}_K^{\dagger}e^{-i\omega_d t}+\hat{s}_Ke^{i\omega_d t})$, with drive strength $\Omega_d$~\cite{2022XiongP245310245310}. Under the strong microwave driving condition, the hybrid system can be described by the Hamiltonian $\hat{H}_{\rm{NKS}}=\Delta_q/2\hat{\sigma}_z+\widetilde{\Delta}_K\hat{s}_K^{\dagger}\hat{s}_K+\bar{\lambda}_{\rm{KS}}(\hat{s}_K\hat{\sigma}_++\hat{s}_K^\dagger\hat{\sigma}_-) -K_d/2(\hat{s}_K^{\dagger 2}+\hat{s}_K^2)$, where $\Delta_q=\omega_q-\omega_d$, $\widetilde{\Delta}_K=\Delta_K-4K\langle\hat{s}_K\rangle^2$, with $\Delta_K=\omega_K-K-\omega_d$, and the enhanced Kerr coefficient is determined by $K_d=2K\langle\hat{s}_K\rangle^2$. Utilizing the Bogoliubov transformation $\hat{m}=\hat{s}_K\cosh r-\hat{s}_K^{\dagger}\sinh r$, with $\tanh(2r)=K_d/\widetilde{\Delta}_K$~\cite{2016LemondeP1133811338,2019BurdP11631165,2021BurdP898902,2021BlaisP2500525005}, the Hamiltonian $\hat{H}_{\rm{NKS}}$ can be expressed as $        \hat{H}_{\rm{NKS}}^{\rm{Sq}}=\Delta_q/2\hat{\sigma}_z+\Delta_K^{\rm{eff}}\hat{m}^{\dagger}\hat{m}+\lambda_{\rm{KS}}^{\rm{eff}}(\hat{m}\hat{\sigma}_++\hat{m}^{\dagger}\hat{\sigma}_-)$, where $\Delta_K^{\rm{eff}}=\widetilde{\Delta}_K/\cosh(2r)$ and $\lambda_{\rm{KS}}^{\rm{eff}}=\bar{\lambda}_{\rm{KS}}\cosh r$. Here we ignore the anti-rotation term $\bar{\lambda}_{\rm{KS}}\sinh r(\hat{m}\hat{\sigma}_-+\hat{m}^{\dagger}\hat{\sigma}_+)$ when condition $\Delta_q,\Delta_K^{\rm{eff}}\gg \bar{\lambda}_{\rm{KS}}\sinh r$ is satisfied. \textit{The coupling strength} of the YIG sphere and the skyrmion qubit \textit{is enhanced exponentially} with the squeezing parameter $r$. The parametric amplification technique can enable strong coupling of skyrmion qubits to much larger YIG spheres with dimensions of tens of micrometers, as illustrated in Fig.~\ref{FIG2}(d).

\begin{figure}
    \centering
    \includegraphics[width=0.48\textwidth]{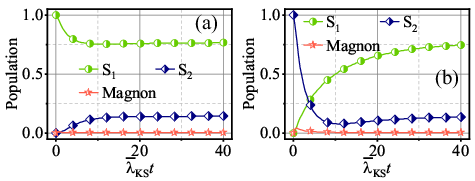}
    \caption{Population conversion dynamics: (a) only qubit 1 in the excited state  and (b)  only qubit 2 in the excited state. $S_1$ and $S_2$ correspond to the population of the first skyrmion and second skyrmion, respectively. The parameters used are $\Delta_{K,1}=\Delta_{q,1}=\Omega_2=\bar{\lambda}_{\rm{KS}}$ and $\gamma_K=10\bar{\lambda}_{\rm{KS}}$.}
    \label{FIG3}
\end{figure}

\textit{Nonreciprocal interactions between skyrmion qubits.}---We now consider a nonreciprocal coupling, mediated by magnons, between two skyrmion qubits. Two qubits are coupled to the same YIG sphere, one of which is driven by microwaves $\hat{H}_{\rm{qd}}=-\Omega_1(e^{i\omega_1 t}\hat{\sigma}_-^1+e^{-i\omega_1 t}\hat{\sigma}_+^1)-\Omega_2(e^{i\omega_2 t}\hat{\sigma}_-^1+e^{-i\omega_2 t}\hat{\sigma}_+^1)$, where $\Omega_{1/2}$ and $\omega_{1/2}$ are the driving strength and frequency, respectively. The system's Hamiltonian, when transformed to the interaction picture of drive $\Omega_1$, is represented by~\cite{SM} $\hat{H}_{\rm{SKSD}}=\Omega_2/2\hat{\sigma}_z^1+\Delta_{q,1}/2\hat{\sigma_z}^2+\Delta_{K,1}\hat{s}_K^\dagger \hat{s}_K +\bar{\lambda}_{\rm{KS}}/2(\hat{s}_K+\hat{s}_K^\dagger)\hat{\sigma}_x^1+\bar{\lambda}_{\rm{KS}}(\hat{s}_K\hat{\sigma}_+^2+\hat{s}_K^\dagger\hat{\sigma}_-^2)$ with $\Delta_{q,1}=\omega_q-\omega_1$ and $\Delta_{K,1}=\omega_K-\omega_1$. In other words, the JC model and the effective Rabi model are combined in a single setup. By taking into account the large dissipation of magnons ($\gamma_K\gg \bar{\lambda}_{\rm{KS}}$) and eliminating the magnon mode adiabatically, the effective master equation is given by $\dot{\hat{\rho}}=-i[\hat{H}_{\rm{coh}},\hat{\rho}]+\Gamma D[\hat{\Sigma}_-]\hat{\rho}$ including the coherent coupling of $\hat{H}_{\rm{coh}}=1/2\mathcal{W}_1\hat{\sigma}_z^1+1/2\mathcal{W}_2\hat{\sigma}_z^2-\mathcal{G}\hat{\sigma}_x^1\hat{\sigma}_x^2$ and the dissipative coupling of $\hat{\Sigma}_-=1/2\hat{\sigma}_x^1+\hat{\sigma}_-^2$. The parameters $\mathcal{W}_1$, $\mathcal{W}_2$, $\mathcal{G}$, and $\Gamma$ are defined in detail in Ref.~\cite{SM}. The system's  quantum Langevin equations (QLEs) can be expressed as
\begin{equation}
    \begin{split}
        \dot{\hat{\sigma}}_-^1&=-\frac{\Gamma}{4}\hat{\sigma}_-^1+\frac{\Gamma}{4}\hat{\sigma}_+^1-\left(-\frac{\Gamma}{4}\hat{\sigma}_-^2+\frac{\Gamma}{4}\hat{\sigma}_+^2\right)\hat{\sigma}_z^1, \\
        \dot{\hat{\sigma}}_-^2&=-\frac{\Gamma}{4}\hat{\sigma}_-^2-\left(-\frac{\Gamma}{4}\hat{\sigma}_-^1-\frac{\Gamma}{4}\hat{\sigma}_+^1\right)\hat{\sigma}_z^2,
    \end{split}
\end{equation}
where we have chosen the parameters as $\mathcal{W}_1=\mathcal{W}_2=\mathcal{G}=0$. The entire QLEs are provided in Ref.~\cite{SM}. The nonlocal damping $\Gamma$ couples the raising and lowering operators of the two qubits; in other words, the nonlocal damping induced by the engineered reservoir mediates a nonlocal damping force on each qubit. Furthermore, the dissipative coupling is asymmetric, allowing for nonreciprocal population conversion. Figures~\ref{FIG3}(a) and (b) depict the dynamics of the system's population conversion when qubits 1 and 2 are excited, respectively. When qubit 1 is excited, it is difficult for qubit 2 to receive the excitation converted by qubit 1, but conversely qubit 1 can easily get the excitation converted by qubit 2.

\begin{figure}
    \centering
    \includegraphics[width=0.48\textwidth]{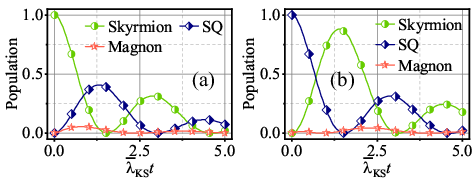}
    \caption{Population conversion dynamics:  (a) for  the case of the skyrmion qubit being excited and (b) the SQ being excited. The parameters used are $\eta=1$, $\phi_e=\pi/2$, $\mathcal{G}_{\rm{SS}}=\bar{\lambda}_{\rm{KS}}$, and $\gamma_K=10\bar{\lambda}_{\rm{KS}}$.}
    \label{FIG4}
\end{figure}

\textit{Nonreciprocal  interactions between skyrmion qubits and SQs.}---We consider a YIG sphere that is coupled to both a skyrmion qubit and a SQ, as described by the Hamiltonian $\hat{H}_{\rm{SKT}}=\omega_q/2\hat{\sigma}_z+\omega_{\rm{Tr}}/2\hat{\sigma}_z^S+\omega_K\hat{s}_K^\dagger\hat{s}_K+\bar{\lambda}_{\rm{KS}}(\hat{s}_K\hat{\sigma}_++\hat{s}_K^\dagger\hat{\sigma}_-)+\mathcal{J}_{\rm{KT}}(\hat{s}_K\hat{\sigma}_+^S+\hat{s}_K^\dagger\hat{\sigma}_-^S)$~\cite{SM}. $\omega_{\rm{Tr}}$ is the SQ's resonance frequency, and $\mathcal{J}_{\rm{KT}}=\mathcal{J}_{\rm{KT}}^0\cos(\omega_{ac}t+\phi_e)$ is the coupling strength between the magnon and the SQ, which can be achieved by tuning the external fluxes~\cite{2022KounalakisP3720537205}. In the interaction picture, the Hamiltonian $\hat{H}_{\rm{SKT}}$ is reduced to $\hat{H}_{\rm{SKT}}=\bar{\lambda}_{\rm{KS}}(\hat{s}_K\hat{\mathfrak{L}}_+ + \hat{s}_K^\dagger\hat{\mathfrak{L}}_-)$, where $\hat{\mathfrak{L}}_-\equiv \hat{\sigma}_-+\eta\hat{\sigma}_-^S e^{i\phi_e}=\hat{\mathfrak{L}}_+^\dagger$ and $\eta=\mathcal{J}_{\rm{KT}}^0/(2\bar{\lambda}_{\rm{KS}})$. The modulation frequency and phase are represented by $\omega_{\rm{ac}}$ and $\phi_e$, respectively. In the large dissipation limit $\gamma_K\gg \bar{\lambda}_{\rm{KS}}, \mathcal{J}_{\rm{KT}}^0$, the magnon modes can be adiabatically eliminated, and the reduced system can be described by the master equation $\dot{\hat{\rho}}=-i[\hat{H}_{\rm{SS}},\hat{\rho}]+\Gamma_{\rm{SS}}D[\hat{\mathfrak{L}}_-]\hat{\rho}$, where $\hat{H}_{\rm{SS}}=\mathcal{G_{\rm{SS}}}(\hat{\sigma}_+\hat{\sigma}_-^S+\hat{\sigma}_-\hat{\sigma}_+^S)$ is the coherent coupling between the skyrmion qubit and the SQ, which can be achieved by an auxiliary cavity. The coherent and dissipative coupling strengths are denoted by $\mathcal{G_{\rm{SS}}}$ and $\Gamma_{\rm{SS}}$, respectively. The system's QLEs can be expressed as
\begin{equation}
    \begin{split}
        \dot{\hat{\sigma}}_-&=-\frac{\Gamma_{\rm{SS}}}{2}\hat{\sigma}_-+\left(i\mathcal{G}_{\rm{SS}}+\frac{\Gamma_{\rm{SS}}}{2}\eta e^{i\phi_e}\right)\hat{\sigma}_-^S\hat{\sigma}_z, \\
        \dot{\hat{\sigma}}_-^S&=-\frac{\Gamma_{\rm{SS}}}{2}\eta^2\hat{\sigma}_-^S+\left(i\mathcal{G}_{\rm{SS}}+\frac{\Gamma_{\rm{SS}}}{2}\eta e^{-i\phi_e}\right)\hat{\sigma}_-\hat{\sigma}_z^S.
    \end{split}
\end{equation}
It is worth noting that the presence of phase $\phi_e$ allows for a mutual balance of coherent and dissipative coupling. This can be used to achieve a nonreciprocal population conversion between two qubits. Figures~\ref{FIG4}(a) and (b) depict the nonreciprocal conversion of population between two qubits in the case of the skyrmion qubit and SQ excitation, respectively. In particular, with $\phi_e=\pi/2$ and $\mathcal{G}_{\rm{SS}}=-\eta\Gamma_{\rm{SS}}/2$, we get $\dot{\hat{\sigma}}_-=-\Gamma_{\rm{SS}}/2\hat{\sigma}_-$ and $\dot{\hat{\sigma}}_-^S=-\Gamma_{\rm{SS}}/2\eta^2\hat{\sigma}_-^S-i\Gamma_{\rm{SS}}\eta\hat{\sigma}_-\hat{\sigma}_z^S$, indicating that the SQ is influenced by the skyrmion qubit, but conversely the skyrmion qubit is unaffected, implying that complete isolation from the SQ to the skyrmion qubit is achieved~\cite{SM}.

\textit{Experimental feasibility.}---Utilizing the topological Hall effect, the Bloch skyrmions in frustrated magnets have been observed experimentally in triangular-lattice \ce{Gd_2PdSi_3}~\cite{2019KurumajiP914918} and breathing-kagom\'e lattice \ce{Gd_3Ru_4Al_{12}}~\cite{2019HirschbergerP58315831}. Additionally, it is expected theoretically that several candidate frustrated magnets will have skyrmions. For instance, frustrated triangular-lattice magnets with the transition-metal ions \ce{NiGa_2S_4}~\cite{2005NakatsujiP16971700,2010StockP3740237402}, \ce{Fe_xNi_{1-x}Br_2} (dihalides)~\cite{1976DayP24812481,1982RegnaultP12831290,1985MooreP2341}, and \ce{\alpha-NaFeO_2}~\cite{2007McQueenP2442024420,2014TeradaP184421184421}, where chemical substitutions can alter the anisotropy of dihalides. Furthermore, \ce{Pb_2VO(PO_4)_2}~\cite{2021UkpongP} with a square lattice is predicted to also have skyrmions. Here the parameters of skyrmions are taken as: the effective spin $\bar{S}=20$, the lattice spacing $a=0.5~\rm{nm}$, the interaction strength $\mathcal{J}_1=1~\rm{meV}$, and the electric polarization $P_E=0.2~\rm{C/m}$ ~\cite{2021PsaroudakiP6720167201,2022PsaroudakiP104422104422}. The anisotropy energy is set to $0.13~\rm{meV}$, and the applied external magnetic field is given as $30~\rm{mT}$~\cite{2021PsaroudakiP6720167201,2022PsaroudakiP104422104422}. Then we can obtain the radius of the skyrmion $\sim 6~\rm{nm}$~\cite{2023XiaP106701106701}. In addition, we suppose that the applied electric field gradient is $570~\rm{V/m}$, which is utilized to control the skyrmion-qubit energy level spacing~\cite{2020YaoP8303283032,2021PsaroudakiP6720167201}. The resonant frequency of the skyrmion qubit constructed in our model is $\omega_q/2\pi\approx 14~\rm{GHz}$.

We assume that the radius of the YIG sphere is $R_K=100~\rm{nm}$, the saturation magnetization is $M_s=587~\rm{kA/m}$, and the distance from the sphere's surface to the skyrmion is $d_K=10~\rm{nm}$~\cite{2023FuwaP6351163511,2021HeiP4370643706,2020GonzalezBallesteroP9360293602,2020GonzalezBallesteroP125404125404,2023HeiP7360273602,2020NeumanP247702247702}. The Kittel mode in the YIG sphere can be excited by an applied bias magnetic field $B_K=500~\rm{mT}$~\cite{2014TabuchiP8360383603,2014ZhangP156401156401}, which ensures that the YIG sphere is in saturation magnetization and is significantly lager than the magnetic field used to stabilize the skyrmion. Additionally, since homogeneous $B_K$ contributes zero energy to skyrmions, the biased magnetic field $B_K$ has no effect on skyrmions' stabilization~\cite{2016LinP6443064430,2021PsaroudakiP6720167201}. Then, the resonant frequency of the magnon is $\omega_K/2\pi\approx14~\rm{GHz}$ and the coupling strength of the skyrmion and the magnon is $\lambda_{\rm{KS}}/2\pi=12.7\rm{MHz}$, which is much smaller than the non-harmonicity of the skyrmion qubit. Assuming that the dissipation of both the magnon and skyrmion qubits is $1~\rm{MHz}$~\cite{2021HeiP4370643706,2023HeiP7360273602}, the hybrid quantum system's cooperativity is $\mathcal{C}\approx 51 \gg 1$. Even though the dissipation of the skyrmion qubit is taken to be $10~\rm{MHz}$, the cooperativity  $\mathcal{C}\approx 5$ is still greater than one, i.e., the system can still reach the strong-coupling regime. For a finite temperature $T=100~{\rm{mK}}$, the equilibrium thermal magnon occupancy numbers are $\bar{n}\approx 0.0012\ll 1$, and the skyrmion quits are also not thermally excited~\cite{2023XiaP106701106701}. In Ref.~\cite{SM}, the feasibility of the proposed scheme here is further analyzed in Sec.~VIII utilizing micromagnetic simulations. In addition, the calculation of the coupling strength in a multilayer structured hybrid system, composed of square or circular  dots and skyrmions, is calculated in detail in Sec.~IX of Ref.~\cite{SM}.

\textit{Conclusion.}---We have proposed a hybrid quantum system composed of YIG micromagnets and skyrmions, and show that it can achieve the strong-coupling regime described by the JC model. We incorporate a microwave drive to make the magnons enter the nonlinear region and then employ parametric amplification techniques to achieve an exponential increase in the coupling strength. The magnon is then used as an intermediary to induce coherent and dissipative couplings between skyrmions or between skyrmions and other quantum systems. Magnon-mediated coherent couplings are utilized to improve the scalability of qubits; together with dissipative couplings, it is possible to achieve a nonreciprocal interaction and response between different qubits.

\begin{acknowledgments}
P.B.L. is supported by the National Natural Science
Foundation of China under Grants No. 12375018 and No. 92065105.
F.N. is supported in part by Nippon
Telegraph and Telephone Corporation (NTT) Research,
Japan Science, and Technology Agency (JST) (via the
Quantum Leap Flagship Program (Q-LEAP), Moonshot
R\&D Grant No. JPMJMS2061, the Asian Office of
Aerospace Research and Development (AOARD) (via
Grant No. FA2386-20-1-4069), and the Foundational
Questions Institute (FQXi) (via Grant No. FQXiIAF19-06).
X.L.H. is Supported by the Fundamental Research Funds for the Central Universities (Program No. xzy022023002).
X.Z. and M.M. acknowledge support by CREST, the Japan Science and Technology Agency (Grant No. JPMJCR20T1).
M.M. also acknowledges support by the Grants-in-Aid for Scientific Research from JSPS KAKENHI (Grants No. JP20H00337 and No. 23H04522), and the Waseda University Grant for Special Research Projects (Grant No. 2023C-140).
\end{acknowledgments}

%

\end{document}